\documentclass[twocolumn]{aastex63}

\usepackage[utf8]{inputenc} 
\usepackage[T1]{fontenc} 

\usepackage{amsmath}
\newcommand{\dif}{\mathrm{d}}
\usepackage{bm}
\usepackage{savesym}
\savesymbol{tablenum}
\usepackage{siunitx}
\restoresymbol{SIX}{tablenum}

\usepackage{booktabs}

\received{2022 February 24}
\revised{2022 April 4}
\accepted{2022 April 5}

\shorttitle{Hurricane-like vortices in protoplanetary disks}
\shortauthors{Gerbig and Laughlin}

\graphicspath{{./}{figures/}}

\newcommand{\new}[1]{{#1}} 

\begin{document}

\title{The Prospects for Hurricane-like Vortices in Protoplanetary Disks}

\correspondingauthor{Konstantin Gerbig}
\email{konstantin.gerbig@yale.edu}

\author[0000-0002-4836-1310]{Konstantin Gerbig}
\affiliation{Department of Astronomy, Yale University, 52 Hillhouse Ave, New Haven, CT 06511, USA}

\author[0000-0002-3253-2621]{Gregory Laughlin}
\affiliation{Department of Astronomy, Yale University, 52 Hillhouse Ave, New Haven, CT 06511, USA}

\begin{abstract}
When ice on the surface of dust grains in protoplanetary disk sublimates, it adds its latent heat of water sublimation to the surrounding flow. Drawing on the analogy provided by tropical cyclones on Earth, we investigate if this energy source is sufficient to sustain or magnify anticyclonic disk vortices that would otherwise fall victim to viscous dissipation. An analytical treatment, supported by exploratory two-dimensional simulations, suggests that even modestly under-saturated flows can extend the lifetime of vortices, potentially to a degree sufficient to aid particle trapping and planetesimal formation. We expect the best conditions for this mechanism to occur will be found near the disk's water ice line if turbulent motions displace gas parcels out of  thermodynamic equilibrium with the dust mid-plane. 
\end{abstract}


\section{Introduction}

Vortices in protoplanetary disks around young stars are often believed to constitute an important phenomenon during planet formation \citep[][]{Adams1995, Barge1995,Godon1999, Manger2018, Flock2020}. In particular, sufficiently long-lived anticyclonic vortices, have been shown capable of attracting and trapping particles, thus providing a potentially favorable environment for dust coagulation and the formation of planetesimals \citep[e.g.,][]{Tanga1996, Klahr1997,Chavanis2000, Lyra2009, Heng2010, Meheut2012, Zhu2014, Raettig2015, Regaly2021}. 

In the consensus theoretical framework, a key location in a protoplanetary disk is the water ice line, which marks the innermost radius at which water vapor and ice can coexist in equilibrium. Depending on the partial pressure of water vapor, the ice line lies at a temperature of $\sim 150$ K \citep[e.g.,][]{Podolak2004, Lecar2006, Martin2012}. At the ice line, if the system is perturbed from phase equilibrium, ice will sublimate to vapor, or conversely, water vapor will deposit onto the silicate surfaces of dust, depending on whether the flow is under- or super-saturated. In the process, the flow is provided with, or deprived of enthalpy given by the latent heat of sublimation of water.

Our goal with this paper is to investigate the possibility that such a thermodynamic disequilibrium between the gas disk and the particle mid-plane can maintain, enhance, and spur the formation of vortices. We draw on the analogy to the well-understood process by which the thermodynamic disequilibrium between the ocean and the atmosphere provides the energy source for tropical cyclones on Earth \citep{Kleinschmidt1951, Emanuel1991}. There, strong surface winds roil the ocean surface and in the process pick up water which \new{increases the moisture content of the previously under-saturated near-surface air.} This exchange transfers enthalpy in form of the water's latent heat of evaporation from the ocean to the atmosphere, and the enthalpy is made accessible as potential energy for thermodynamic transformation. In particular, \new{once the flow has acquired enough H$_2$O to achieve saturation,}  condensation \new{and cloud formation} converts latent heat to sensible heat which in turn drives vigorous upwards convection in the eye-wall of the hurricane. This, in turn, decreases surface-level air pressure and accelerates surface-level air into vortical motion via Coriolis deflection around the storm's pressure minimum. The result is strong azimuthal circulation due to conservation of angular momentum, and likewise a transverse circulation due the mass conservation, in which the surface level inflow is balanced by a radial outflow at the tropopause. At this high-altitude ($h\sim10-15\,$km) level, the atmosphere becomes optically thin in the infrared and the flow can radiate excess heat. Positive feedback is achieved due to faster wind speeds increasing the rate of enthalpy transfer from ocean to atmosphere, thus intensifying convection, ultimately leading to a mature hurricane, which is maintained as long as the hurricane is over the ocean and can draw in sufficiently under-saturated air. In steady-state, the frictional losses, namely dissipation at the ocean-atmosphere interface and radiative losses at the tropopause, are in balance with the latent heat flux.

A similar disequilibrium between ice-coated dust grains in a mid-plane layer and surrounding gas flow in protoplanetary disks will be associated with a source of enthalpy that scales with the level of under-saturation. Our goal is to carry out an initial investigation of the prospects for this mechanism to operate in protoplanetary disks. We model hurricane-like heat engines in two-dimensional hydrodynamic simulations by parameterizing both the thermodynamic disequilibrium and the mixing process. 

The paper is structured as follows. Sect.~\ref{sect:theory} outlines the physical model of latent heat flux driven vortices, applied to protoplanetary disks. In Sect.~\ref{sect:nummethods} we present our numerical implementation, while the obtained results are in Sect.~\ref{sect:numresults}. Finally, we discuss our work in Sect.~\ref{sect:discussion}

\section{Vortex model}

\label{sect:theory}

The foregoing discussion suggests that the structure and the thermodynamical properties of protostellar disks allow one to draw strong physical analogies with terrestrial hurricanes. In this section we elaborate the specific details of a model that is based on this analogy. 

\subsection{Latent heat flux}

When a vortex near the ice line of a protoplanetary disk mixes with icy dust grains from the mid-plane, any under-saturation will lead to sublimation which increases the flow's water vapor content, and thus is associated with a net enthalpy transport from dust at the mid-plane to the gas flow that is proportional to the specific latent heat of water sublimation $L_\mathrm{s}$. In analogy to Hurricanes, we characterize the bulk latent heat flux LHF (units erg/s) over a surface $\dif A$ as \citep[compare to e.g.,][]{Liu2011}
\begin{align}
\label{eq:LHF_first}
   \mathrm{LHF} = \frac{\dif(L_\mathrm{s} q)}{\dif t} = L_\mathrm{s} C_\mathrm{E}\rho |\bm{v}| \Delta q \dif A,
\end{align}
where $|\bm{v}|$ is the flow's wind speed\new{, and  $\rho$ is the total gas density.} $C_\mathrm{E}$ is the dimensionless turbulent exchange coefficient which describes how efficient humidity is transferred to the flow. Note that the turbulent exchange coefficient itself is believed to depend on flow velocity \citep[e.g.,][]{Cronin2019}. We discuss this further in Sect.~\ref{sect:mixing_model}. $\Delta q$ \new{is the under-saturation of the flow} given by
\begin{align}
\label{eq:under-saturation}
   \Delta q = q_* - q,
\end{align}
where $q$ is the vapor mixing ratio (also known as specific humidity) of the under-saturated flow that mixes with the dust-layer. It is defined as the ratio of vapor density (absolute humidity) $\rho_\mathrm{v}$ over total gas density $\rho$, i.e.
\begin{align}
\label{eq:mixingratio}
    q = \frac{\rho_\mathrm{v}}{\rho} = \frac{\rho_\mathrm{v}}{\rho_\mathrm{d} + \rho_\mathrm{v}},
\end{align}
where $\rho_\mathrm{d}$ is the density of ``dry'' gas. In Eq.~\eqref{eq:under-saturation}, $q_*$ is the vapor mixing ratio at saturation.

\subsection{Phase equilibrium and saturation}

\begin{figure*}
    \centering
    \includegraphics[width = \textwidth]{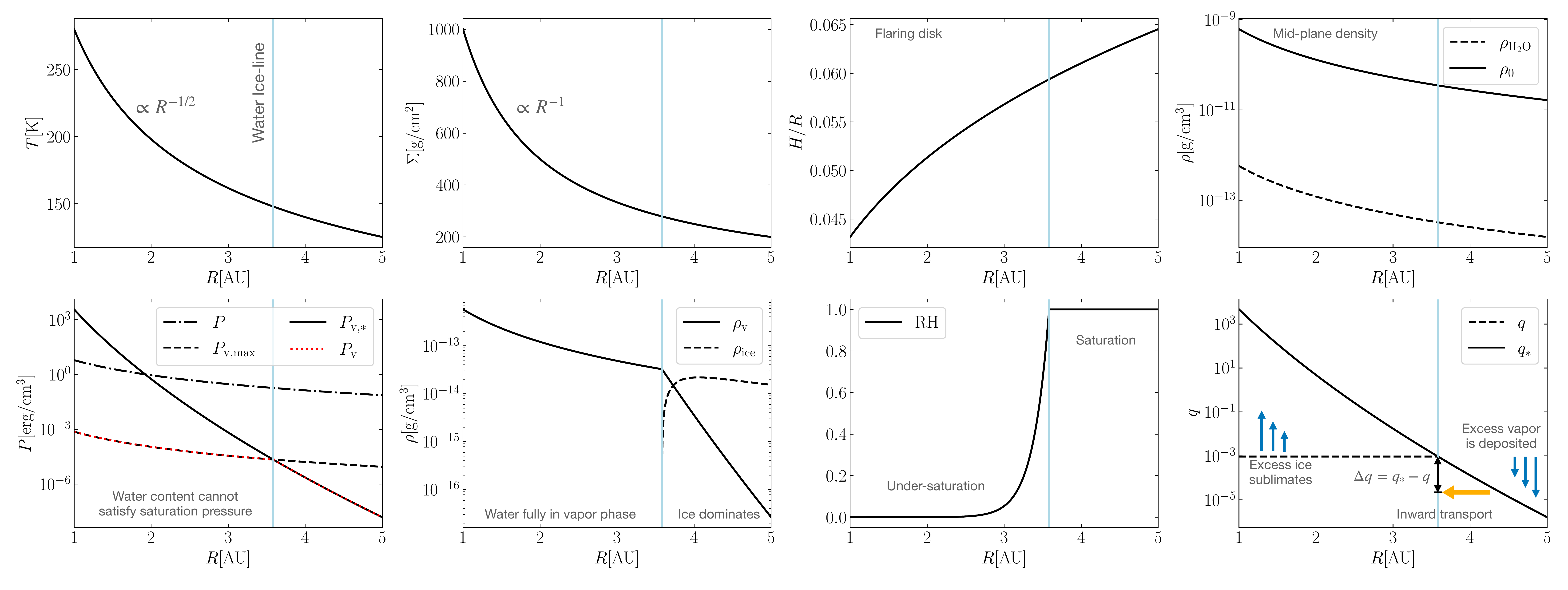}
    \caption{Visualization of key properties of a steady-state disk between 1 AU and 5 AU. Our fiducial model places the water ice line at $\sim 3.6$ AU (blue line). Shown are as a function of distance to star $R$, from left to right, top row: temperature profile $T$ and surface density profile $\Sigma$ based on the MMSN \citep[][]{Hayashi1981}, disk aspect ratio $H/R$, mid-plane gas density $\rho_0$ and (total) water density using a fiducial H$_2$O abundance of $1.2 \cdot 10^{-4}$ \citep[][]{Furuya2013}. Bottom row: total pressure $P$ and partial vapor pressure $P_\mathrm{v}$ which is the minimum of saturation vapor pressure $P_\mathrm{v,*}$ and maximum  partial pressure $P_\mathrm{v, max} = \rho_{\mathrm{H}_2\mathrm{O}}\mathcal{R}_\mathrm{v}T$, water vapor and ice density, relative humidity RH \citep[compare to][]{Ros2019}, and lastly vapor mixing ratio at saturation $q_*$ and actual mixing ratio $q$. We argue that an inwardly transported fluid parcel would be under-saturated with relative to its new environment.
    }
    \label{fig:sat_ratio}
\end{figure*}

Phase equilibrium between water ice and vapor is characterized by the relative humidity 
\begin{align}
    \mathrm{RH} = \frac{P_\mathrm{v}}{P_\mathrm{v,*}},
\end{align}
also known as saturation ratio, being equal to unity, i.e. $\mathrm{RH} = 1$. Here, $P_\mathrm{v}$ is the pressure of the water vapor phase, for an ideal gas given by $P_\mathrm{v} = \rho_\mathrm{v} \mathcal{R}_\mathrm{v} T$. It is thus a partial pressure of the total gas pressure $P = \rho \mathcal{R}_\mathrm{g} T$, where $\mathcal{R}_\mathrm{v} = \mathcal{R}/M_{\mathrm{H}_2\mathrm{O}} = 4.615 \cdot 10^6$ ergs/g/K, and $\mathcal{R}_\mathrm{g} = \mathcal{R}/M_\mathrm{g} = k_\mathrm{B}/(\overline{\mu} m_\mathrm{p})= 3.568 \cdot 10^7$ ergs/g/K are specific gas constants for water vapor and the total gas respectively, with mean molecular weight of the gas $\overline{\mu} = 2.33$ (assuming solar composition), molar mass of water $M_{\mathrm{H}_2\mathrm{O}} = 18.015$ g/mol, proton mass $m_\mathrm{p}$, universal gas constant $\mathcal{R}$, and Boltzmann constant $k_\mathrm{B}$. $P_\mathrm{v,*}$ is the saturation vapor pressure. We can deduce $P_\mathrm{v,*}$ from the curve of vapor-ice coexistence for water, the slope of which is given by the Clausius-Clapeyron relation \citep[][]{Clausius1850}
\begin{align}
\label{eq:CC_relation}
    \frac{\dif P_\mathrm{v}}{\dif T} = \frac{L_\mathrm{s}}{T \Delta v},
\end{align}
where $\Delta v$ is the change in specific volume associated with the phase transition, and $L_\mathrm{s}$ is the latent heat of water sublimation. Following \citet{Ros2019}, we use
\begin{align}
\label{eq:saturation_vapor_pressure}
    P_\mathrm{v,*} =P_\mathrm{v,*,0}\exp\left(-\frac{T_{*,0}}{T}\right),
\end{align}
where $P_\mathrm{v,*,0} = 6.034 \cdot 10^{12}$ g/cm/s$^2$ and $T_{*,0} = 5938$ K are material constants of water \citep{Haynes1992}. This implies for the saturation vapor mixing ratio, again using an ideal equation of state,
\begin{align}
    q_\mathrm{*} = \frac{P_\mathrm{v,*}}{\rho\mathcal{R}_\mathrm{v}T} = \frac{q}{\mathrm{RH}} = \frac{\Delta q}{1- \mathrm{RH}}.
\end{align}
It is important to highlight that $P_\mathrm{v,*}$ and $q_\mathrm{*}$ are not physical properties of the system state, but only quantify partial pressure $P$ and mixing ratio $q$ respectively necessary to achieve phase equilibrium.

The system can only establish equilibrium, however, if there is enough water available to satisfy $P_\mathrm{v} = P_\mathrm{v,*}$. Thus, the (water) iceline can be defined as the furthest-in location where the saturation criterion can be met. Outside the iceline, excess vapor will be deposited onto dust grains, such that $P_\mathrm{v} = P_\mathrm{v,*} \Leftrightarrow \mathrm{RH} = 1$ to maintain phase equilibirum \citep[][]{Ros2019}. Inside the ice line, all the available H$_2$O is in the vapor phase, yet the resulting partial pressure is less than the required $P_\mathrm{v,*}$, resulting in a phase disequilibrium, in particular in an under-saturated flow. 

The iceline location  depends on temperature and density profiles and the water abundance. We proceed with a fiducial disk model where temperature scales as
\begin{align}
\label{eq:Hayashi_temp}
    T = T_0 \left(\frac{R}{R_0}\right)^{-\beta_\mathrm{T}}\, ,
\end{align}
and the gas surface density scales as
\begin{align}
\label{eq:Hayashi_sd}
    \Sigma = \Sigma_0 \left(\frac{R}{R_0}\right)^{-\beta_\Sigma},
\end{align}
with distance to star $R$. For the MMSN \citep[][]{Hayashi1981}, $R_0 = 1$ AU, $T_0$ = 280 K, $\beta_\mathrm{T} = 1/2$, $\Sigma_0 = 1000$ g/cm$^2$, and $\beta_\mathrm{\Sigma} = 1$. Vertical hydrostatic equilibrium implies that the mid-plane gas density is $\rho_0 = \Sigma/(\sqrt{2\pi}H)$. $H = c_\mathrm{s}/\Omega$ is the gas pressure scale height given by the ratio of local speed of sound $c_\mathrm{s} =\sqrt{\gamma k_\mathrm{B}T/(\overline{\mu}m_\mathrm{p})}$  and Keplerian angular frequency $\Omega = \sqrt{GM_\odot/R^3}$. Here, we introduced adiabatic index $\gamma = 5/3$, gravitational constant $G$ and solar mass $M_\odot$. With these power law scalings, we can write the vapor mixing ratio at saturation $q_*$ as a function of $R$ as
\begin{align}
\label{eq:sat_mixing_ratio}
    q_*(R) &= C_{q_*} \left(\frac{R}{R_0}\right)^{\beta_\Sigma + \frac{\beta_T}{2} + \frac{3}{2}}e^{-\left(\frac{T_{*,0}}{T_0}\right)\left(\frac{R}{R_0}\right)^{\beta_\mathrm{T}}}, \\
    C_{q_*} &= \frac{P_\mathrm{v,*,0}}{\Sigma_0 \mathcal{R}_\mathrm{v}T_0}\sqrt{\frac{2\pi k_\mathrm{B}T_0 R_0^3}{\overline{\mu}m_\mathrm{p}G M_\odot}}.
\end{align}

For the water abundance, we assume a fiducial value of $n_{\mathrm{H}_2\mathrm{O}}/ n_{\mathrm{H}} = 1.2 \cdot 10^{-4}$ \citep[][]{Furuya2013}, which implies for the total (vapor + ice) water density $\rho_{\mathrm{H}_2\mathrm{O}} = (n_{\mathrm{H}_2\mathrm{O}}/ n_{\mathrm{H}}) (M_{\mathrm{H}_2\mathrm{O}}/M_\mathrm{H}) \rho_0$. The vapor pressure is therefore
\begin{align}
    P_\mathrm{v} = \mathrm{min}\left( \rho_{\mathrm{H}_2\mathrm{O}}\mathcal{R}_\mathrm{v}T,   P_\mathrm{v,*}\right),
\end{align}
with the iceline located where the two arguments of the minimum function are equal. The water vapor and ice densities are then just $\rho_\mathrm{v} = P_\mathrm{v}/(\mathcal{R}_\mathrm{v}T)$ and $\rho_\mathrm{ice} = \rho_{\mathrm{H}_2\mathrm{O}} - \rho_\mathrm{v}$. An alternate approach to calculating ice and vapor densities is to assume an ice fraction in dust grains \citep[of order 0.5 in models by][]{Min2011}, and a dust-to-gas ratio \citep[of order $10^{-2}$ in the ISM,][]{Savage1972}. This leads to a water abundance of $n_{\mathrm{H}_2\mathrm{O}}/ n_{\mathrm{H}} \sim 2\cdot 10^{-4}$ that is consistent with our chosen value.

Fig.~\ref{fig:sat_ratio} shows the radial variations of the properties introduced in this section. For the utilized MMSN disk model, the iceline is located at $R_\mathrm{0} \sim 3.6$ AU, which is somewhat further out than what is obtained in more sophisticated considerations of the temperature profile \citep[e.g,][]{Kennedy2008, Martin2012}.

\new{We note, that the microphysics that govern H$_2$O deposition and sublimation in disks are not well constrained \citep[for a review see e.g.,][]{Henning2013}. We assume that the associated timescales are short compared to gas transport.} \new{In addition}, our model neglects chemical reactions. For example, UV photons can photodissociatie H$_2$O ice \citep[][]{Gerakines1996}, in particular if dust grains are lifted to the disk surface via turbulent mixing \citep[][]{Furuya2013}. Likewise, water formation through, e.g. hot neutral chemistry \citep[e.g.,][]{Du2014} is not considered. As a result, in our model, the total water density can only change via radial advection. 

\subsection{Phase disequilibria due to radial mixing}

Indeed, any such radial transport of gas parcel, for example epicyclic oscillations or turbulent motions, introduces thermodynamic disequilibrium. In particular, if a fluid parcel at $R_1$ in phase equilibrium with vapor mixing ratio $q_1 = q_*(R_1)$ were to displaced radially inward to radius $R_2 < R_1$, it would be under-saturated relative to its new environment where the ambient saturation mixing ratio is $q_{*}(R_2) > q_1$. The parcel would move towards phase equilibrium by sublimating water ice, which would increase the flow's enthalpy in form of latent heat of water sublimation to the flow according to Eq.~\eqref{eq:LHF_first}. Since the amount of available ice --- a requirement for any latent heat transfer --- increases outwards, but the gradient in saturation vapor pressure increases inwards (see Fig.~\ref{fig:sat_ratio}), the location just outside the water iceline constitutes a sweet spot where the achieved latent heat flux is maximal. For example for our fiducial model, a radial dispacement of one gas scale height leads to $\Delta q \sim 10^{-4}$ just outside the ice line, but $\Delta q \sim 10^{-14}$ at 10 AU which is negligible.

In our analysis, we remain largely agnostic to the origin of radial displacements causing disequilibria, and treat $\Delta q$ as a free parameter and will investigate values of $10^{-6} \leq \Delta q \leq 10^{-4}$.

\vspace{1cm}

\subsection{Vortex energetics}

The latent heat flux increases the water vapor mixing ratio and thus the total enthalpy of the flow 
\begin{align}
    h = C_P T + L_\mathrm{s} q,
\end{align}
where $C_\mathrm{P} = \dif h / \dif T$ is the specific heat capacity at constant pressure, and we assumed the ice content to be neglible. The first law of thermodynamics for a canonical ensemble $\dif h = T\dif s + 1/\rho \dif P$, \new{with specific entropy $s$,} becomes
\begin{align}
\label{eq:firstlaw}
    T\dif s = C_P\dif T + \dif (L_\mathrm{s}q) -\frac{1}{\rho}\dif P.
\end{align}
Using Bernoulli's equation, the full energy equation is given by \citep[][]{Emanuel1991}
\begin{align}
\label{eq:energy_balance}
    \dif \left(\frac{1}{2}|\bm{u}|^2 + \nabla \Phi  + C_P T + L_\mathrm{s}q\right) - T\dif s + F\dif l = 0.
\end{align}
The first term represents the total internal energy of the flow including from left to right the kinetic energy, the gravitational potential energy, the sensible heat, and the latent heat stored in water vapor. The second term is the heat input that must be provided by the ice nuclei, and the third term is any frictional dissipation of energy along the flow path $\dif l$. This may entail shedding of Rossby waves, and any turbulent dissipation induced by (for example) Keplerian shear. We constrain our first-step investigation to optically thick disks, which are primarily heated through stellar irradiation rather than accretion heating \citep[][]{Armitage2015}. As such, the disk vortices that we model do not experience radiative losses.

In a steady-state, the flow's internal energy is constant, and integration of Eq.~\eqref{eq:energy_balance} implies that along any stream line, heating is balanced by friction
\begin{align}
\label{eq:steadystate}
    \int T\dif s = \int F \dif l.
\end{align}
The heating is provided by the enthalpy flux associated with the change in vapor mixing ratio as the flow picks up water vapor \citep[compare with][]{Emanuel1991}. Using the  first law in Eq.~\eqref{eq:firstlaw}, the heating can be written
\begin{align}
    T\dif s = \mathcal{R} T \dif \ln P + \dif (L_\mathrm{s} q).
\end{align}
Here, we have assumed that while the flow itself is adiabatic, its mixing with the dust layer proceeds isothermally. In other words, the disequilibrium is characterized by phase imbalance alone, not by a temperature gradient \new{between ice grains and gas flow}.

Eq.~\eqref{eq:steadystate} implies that along a streamline
\begin{align}
    \mathcal{R}T \int \dif \ln P + \int  \dif (L_\mathrm{s} q) = \int F \dif l
\end{align}
and thus, if the streamline is closed,
\begin{align}
    \oint T \dif s = \oint \dif (L_\mathrm{s} q) = L_\mathrm{s}  C_\mathrm{E} \oint \Delta q \rho |\bm{v}|  \dif A \dif t.
\end{align}
In a steady-state, the energy dissipation rate of the vortex is exactly balanced by the latent heat flux, i.e.
\begin{align}
\label{eq:dissipationrate_prediction}
    \frac{\dif}{\dif t} \oint F \dif l =  \mathrm{LHF}_\mathrm{tot} = L_\mathrm{s} C_\mathrm{E} \rho \Delta q |\bm{v}| A,
\end{align}
where $A$ is the total area of contact between vortex flow and ice layer.

In order for this mechanism to operate, the acquired latent heat of sublimation ---  an internal energy --- must be converted into work. We model this ``piston'' in an ad-hoc manner by prescribing an increase in the flow momentum, such that $\partial\bm{v}^2/ \partial t \propto \mathrm{LHF}$. This is in direct analogy to terrestrial hurricanes, where latent heat-enhanced convection locally increases the atmospheric scale height, thereby lowering the ambient pressure and advecting in more air. We draw further attention to Sect.~\ref{sect:3dmodel}, where we discuss a proposed three-dimensional structure of this mechanism in protoplanetary disks that includes convection.

\subsection{Summary of model assumptions}

It is worth reiterating the assumptions that inform Eq.~\eqref{eq:dissipationrate_prediction}. 
\begin{itemize}
    \item The enthalpy transfer from the dust mid-plane to gas flow is driven by the thermodynamic disequilibrium associated with an under-saturation of the gas given by $\Delta q$. The efficiency of the latent heat transfer characterized by constant turbulent exchange coefficient $C_\mathrm{E}$ (see Sect.~\ref{sect:mixing_model}). 
    \item The thermodynamic disequilibrium, specifically the under-saturation $\Delta q$ can be maintained. For closed streamlines, this requires the transport of material to regions where deposition and de-saturation occurs, i.e. to greater $R$ where $q_*$ is smaller. For open systems, this can be achieved if the vortex can draw-in new under-saturated gas \citep[compare to terrestrial hurricane in][]{Emanuel1991}. 
    \item The dust-midplane acts as a heat bath that keeps the flow at constant temperature as it mixes with the flow. The mixing process is therefore isothermal.
    \item Along a closed streamline, $\oint \dif \ln P = 0$, i.e. any expansion is followed by compression of equal magnitude. A special case satisfying this assumption is a flow, in which heat input is exclusively associated with an increase in internal energy and not with any work (or temperature), i.e. an isothermal and isobaric phase transition. The increase in internal energy is associated with an increase in flow speed to achieve positive feedback.
\end{itemize}

\section{Numerical Setup}
\label{sect:nummethods}

We employ Athena++\footnote{\href{https://github.com/PrincetonUniversity/athena}{https://github.com/PrincetonUniversity/athena}} \citep[][]{Stone2020} which solves hydrodynamic equations using a high-order Godunov scheme. Hereby, the hydrodynamic equations are solved on a Eulerian grid, and in the shearing sheet reference frame \citep[][]{Stone2010}. The $x$, and $y$ coordinates correspond to radial and azimuthal directions respectively. We utilize an adiabatic equation of state. As a first step, we perform two-dimensional shearing sheet simulations with a fiducial $N_x \times N_y = 256 \times 256$ mesh. In Sect.~\ref{sect:vortex_dissipation} we extend the calculations to explore variations in the grid resolution. The physical dimensions are $L_x = L_y = 2H$.

The full system of equation reads
\begin{align}
    \frac{\partial \rho}{\partial t} + \nabla \cdot (\rho \bm{v}) &= 0, \\
    \begin{split}
        \frac{\partial(\rho\bm{v})}{\partial t} + \nabla \cdot (\rho \bm{v}\bm{v} + P) &\\ - 3 \rho \Omega^2 x \bm{\hat{x}} + 2 \Omega \bm{\hat{z}} \times \rho \bm{v} + \bm{f}_\mathrm{heat}- \bm{f}_\mathrm{damp} &=0 ,
    \end{split}\\
    \frac{\partial E}{\partial t} + \nabla \cdot (E+P)\bm{v} - 3 \Omega^2 \rho \bm{v} \cdot x \bm{\hat{x}} &= 0,
\end{align}
where 
\begin{align}
    E = \frac{P}{\gamma -1} + \frac{1}{2}\rho \bm{v}\cdot \bm{v}.
\end{align}
Terms containing the Keplerian angular frequency $\Omega$ are associated with the shearing sheet and include the Coriolis and centrifugal forces, as well as the linearized background shear $- 3/2 \Omega x$. In the momentum equation, $\bm{f}_\mathrm{damp}$ is a damping coefficient associated with soak zones, and $\bm{f}_\mathrm{heat}$ is the driving coefficient corresponding to the heat-work conversion outlined in the previous section. We construct $\bm{f}_\mathrm{heat}$ to distribute the momentum provided by the enthalpy of sublimation such that $v_x/v_y = \Delta v_x/\Delta v_y$. Thus in each zone, there is an increase in specific momentum given by
\begin{align}
    \rho \Delta v_\mathrm{x} &= \rho\sqrt{\frac{2\mathrm{LH}}{1+\left(\frac{v_x}{v_y}\right)^2}}\, , \\
    \rho \Delta v_\mathrm{y} &= \rho\sqrt{\frac{2\mathrm{LH}}{1+\left(\frac{v_y}{v_x}\right)^2}}\, ,
\end{align}
where $\mathrm{LH} = \mathrm{LHF}\cdot \dif t = L_\mathrm{s} C_\mathrm{E} \rho |\bm{v}| \Delta q \dif A \dif t$ is the specific latent heat acquired in a given time step $\dif t$ within area $\dif A$. In each zone, we set $\dif A = L_x/N_x \cdot L_y/N_y = 6.1 \cdot 10^{-5} H^2$ as the domain of contact with the dust mid-plane. Deviations of the boundary layer area from this simple shape are absorbed in the turbulent exchange coefficient, $C_\mathrm{E}$.

\subsection{Scaling Relations}

In code units, we set $\Omega^{-1}$, $c_\mathrm{s}$, the unperturbed background density $\rho_0$, and the unperturbed temperature $T_0$ to unity, which implies that the gas scale height, $H$, the unperturbed pressure, $P_0$, and the reference volume, $V = H^3$ are also unity in code units. Using Eq.~\eqref{eq:Hayashi_temp}, and $\mu = 2.33$ the sound speed can be expressed in cgs units with
\begin{align}
    c_\mathrm{s} = 9.96 \cdot 10^4 \frac{\mathrm{cm}}{\mathrm{s}} \left(\frac{R}{1 AU}\right)^{-1/4}.
\end{align}
The energy per unit mass corresponding to a value of unity in code units is given by $c_\mathrm{s}^2/2 = 4.96 \cdot 10^9  ({R}/{1 AU})^{-1/2}\, {\mathrm{erg}}/{\mathrm{g}}$.
We can therefore express the latent heat of water sublimation, $L_\mathrm{s} = 3 \cdot 10^{10}\,$ erg/g \citep{Datt2011} in terms of code units as
\begin{align}
    L_\mathrm{s} = 6.05 \cdot \left(\frac{R}{\mathrm{AU}}\right)^{1/2}.
\end{align}

\subsection{Vortex initial conditions}

\label{sect:vortex_initial_conditions}

\begin{figure}
    \centering
    \includegraphics[width = \linewidth]{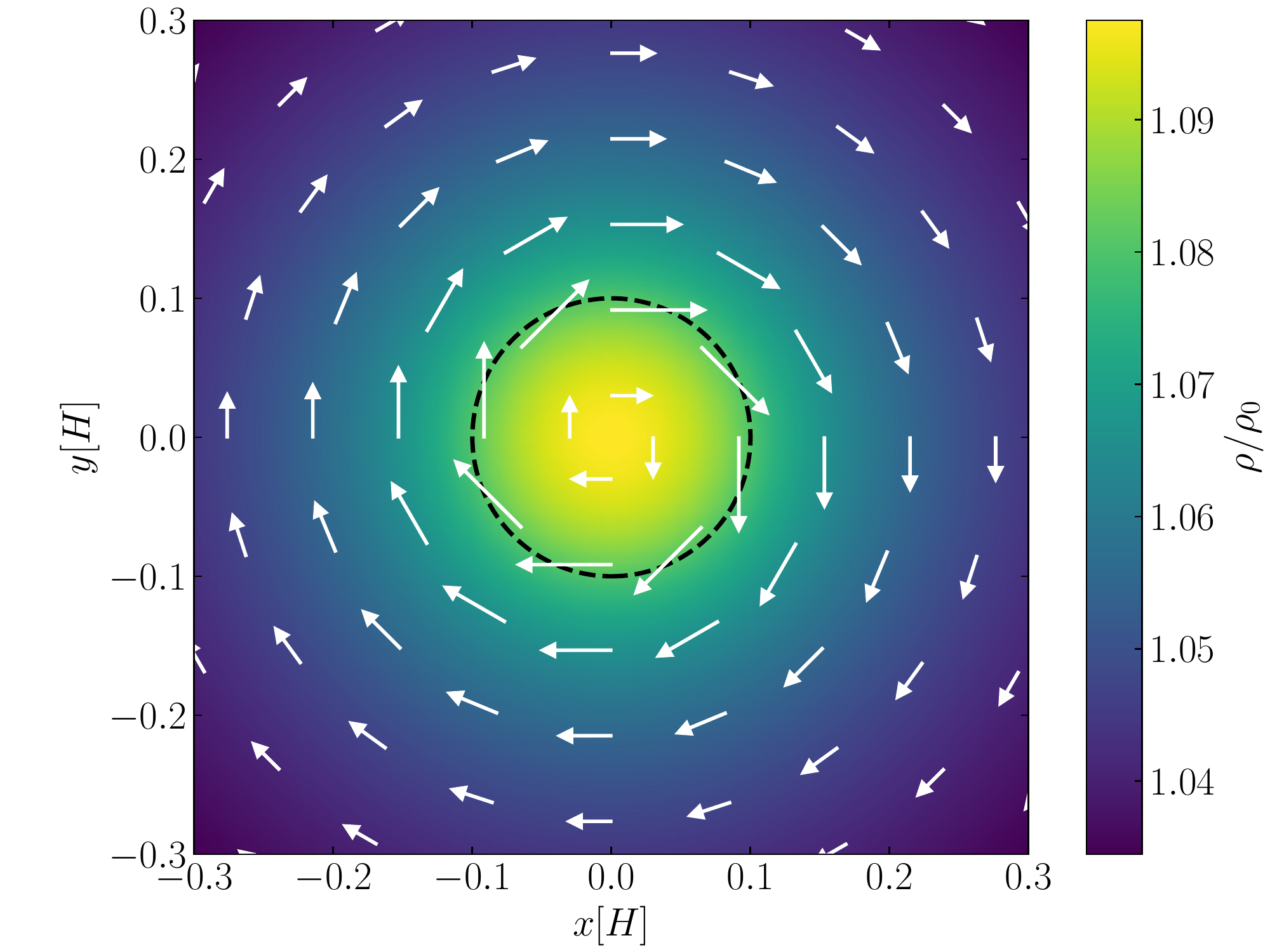}
    \caption{Vortex initial conditions after \citet{Adams1995} for $\Gamma = 0.1$. Color corresponds to density according to Eq.~\eqref{eq:init_dens}, and arrows represent velocity after Eq.~\eqref{eq:init_vel} in arbitrary units. The black circle marks the characteristic radius $R_\mathrm{c} = 0.1 H$, where circular velocities peak, and within which the vortensity is constant.}
    \label{fig:vortex_initcond}
\end{figure}

\citet{Adams1995} provide circular symmetric vortex solutions for a vortex in geostrophic balance in the limit of zero shear. Similarly to \citet[][]{Seligman2017}, we seed our simulation with an extended vortex which is characterized by a vortex strength (circulation) of $\Gamma$ (units cm$^2$/s) and a characteristic radius $R_\mathrm{c}$, within which the vortensity (vorticity over surface density) is set constant. If $\varpi$ is the radial coordinate, and $k_\mathrm{R} = \Omega/c_\mathrm{s} = 1/H$ the Rossby wavenumber, the velocity profile is given by \citep[][]{Adams1995}
\begin{align}
\label{eq:init_vel}
    v_\varphi(\varpi) = \frac{\Gamma}{\pi R_\mathrm{c}}\begin{cases}
    K_1(k_R R_\mathrm{c})I_1(k_R\varpi) &, \varpi \leq R_\mathrm{c} \\
    I_1(k_R R_\mathrm{c})K_1(k_R\varpi) &, \varpi > R_\mathrm{c}
    \end{cases},
\end{align}
where $I_\alpha$ and $K_\alpha$ are modified Bessel functions of the first and second kind respectively. The corresponding density profile that in the absence of shear would establish geostrophic balance is
\begin{align}
\label{eq:init_dens}
    \rho(\varpi) =\rho_0\cdot \exp\left(-\frac{2\Omega}{c_\mathrm{s}^2}\Phi\right),
\end{align}
with stream function
\begin{align}
    \Phi = \frac{\Gamma}{\pi k_R R_\mathrm{c}}\begin{cases}
    K_1(k_R R_\mathrm{c})I_0(k_R \varpi) + C_\Phi
     &, \varpi \leq R_\mathrm{c}
     \\
    - I_1(k_R R_\mathrm{c})K_0(k_R \varpi) &, \varpi > R_\mathrm{c}
    \end{cases},
\end{align}
and $C_\Phi = -K_1(k_R R_\mathrm{c})I_0(k_R R_\mathrm{c})- I_1(k_R R_\mathrm{c})K_0(k_R R_\mathrm{c})$. Density and velocity profiles for $\Gamma = 0.1$, and $R_\mathrm{c} = 0.1 H$ are shown in Fig.~\ref{fig:vortex_initcond}. Azimuthal velocities are maximal at $\varpi = R_\mathrm{c}$, which is also the location of the steepest density gradient.

The solution of \citet{Adams1995} is stable in the absence of shear. Simulations by \citet{Godon2000, Seligman2017} indicate that in the presence of background shear, the vortex adjusts, sheds Rossby waves, and is forced into an oval shape. We discuss this readjustment further in Sect.~\ref{sect:vortex_dissipation}. Work by \citet{Lyra2021} has recently presented disk vortex solutions in elliptical coordinates, but an analytical solution that incorporates an equilibrium vortex into a background Keplerian shear has not yet been found. \new{Indeed, the elliptical instability \citep[][]{Lesur2009} caused by resonance of vortex turnover time and epicyclic frequency is expected to trigger for all two-dimensional vortices in disks, suggesting a physical equilibrium solution may not exist.}

\subsection{Boundary conditions and soak zones}

We adopt boundary conditions that are periodic in $y$ and shear-periodic in $x$. In the absence of a latent heat flux, we expect seeded vortices to dissipate over time due to shedding of Rossby waves and friction with the shear flow generated by numerical viscosity inherent to the computational method. (We do not include an explicit viscous dissipation.). In order to prevent Rossby waves from propagating across the shear-periodic boundaries, reappearing in the domain, and interfering with the vortex, we dedicate the 5\% of the grid cells closest to the radial simulation boundaries as soak zones. These zones act to remove momentum from the simulation, characterized by $\bm{f}_\mathrm{damp}$, thereby improving the isolation of the vortex.

\subsection{Diagnostic measures}

As a measures of vortex dissipation, we take the maximum value of the flow's vorticity
\begin{align}
    \omega = \left(\nabla \times \bm{v}\right)_z = \frac{\partial v_y}{\partial x} - \frac{\partial v_x}{\partial y},
\end{align}
as well as total kinetic energy $E_\mathrm{kin}$. We expect our vorticities to decay exponentially \citep[compare to e.g.,][]{Godon1999}. Hence, we define the exponential decay time $\tau$ via
\begin{align}
\label{eq:decaytime}
    \omega(t) = \omega_0 \exp\left(-\frac{t}{t_\mathrm{d}}\right) + \omega_\mathrm{eq},
\end{align}
where $\omega_0$ is the vortex's initial maximum vorticity and $\omega_\mathrm{eq}$ some equilibrium vorticity.

\section{Numerical Results}
\label{sect:numresults}

\begin{figure*}[t]
    \centering
    \includegraphics[width = \textwidth]{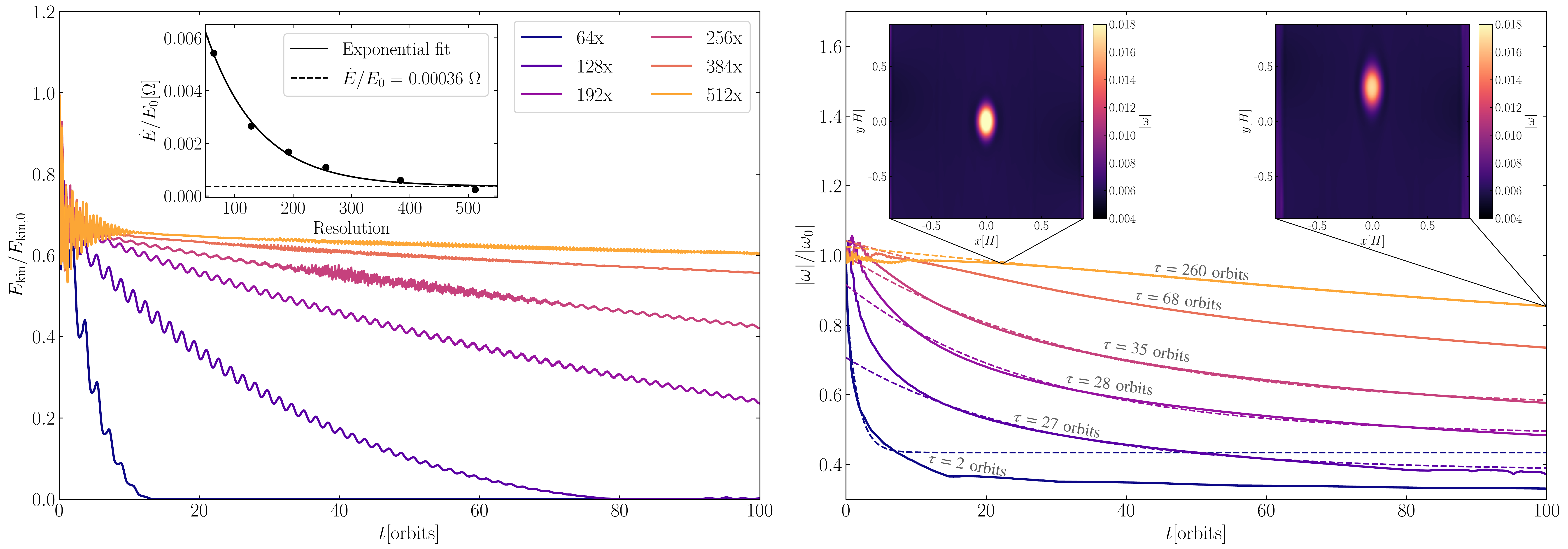}
    \caption{Rate of vortex decay tracked by total kinetic energy (left panel) and maximum vorticity (right panel). The colors correspond to different resolutions.  The energy dissipation rate decreases exponentially with resolution as the numerical viscosity decreases (Sub-plot in the left panel panel). We calculated the decay time $\tau$ in Eq.\eqref{eq:decaytime} by fitting exponential functions to the vorticity profiles. The right panel includes two snapshots of the vorticity distribution in the highest resolution ($512 \times 512$) simulation. }
    \label{fig:dissipation}
\end{figure*}

\subsection{Vortex adjustment and dissipation}

\label{sect:vortex_dissipation}

Before including the heat engine implied by Eq.~\eqref{eq:LHF_first}, we test our numerical setup by evolving the vortex initial condition \new{(with $\Gamma = 0.1$, $R_\mathrm{c} = 0.1 H$)} for a variety of different resolutions for 100 orbits and measure the ensuing vortex dissipation rate. Evolution of vorticity and total kinetic energy as a function of grid resolution is shown in Fig.~\ref{fig:dissipation}. The lowest resolution simulation cannot maintain the vortex. Geostrophic balance is never established and the vortex effectively vanishes within 10 orbits. (Note that there is some vorticity even in the absence of the vortex due to the shear flow.) The other simulations maintain a vortex over the course of 100 orbits. All vortices lose kinetic energy and vorticity, the rate of which decreases smoothly with increasing resolution. The right panel of Fig.~\ref{fig:dissipation} includes two snapshots the vorticity of the highest-resolution simulation $512 \times 512$. Dissipation is manifested in drag (decrease in maximum vorticity) and diffusion (increase in radial extent).

The resolution also impacts the adjustment time required for the vortex to adapt to the Keplerian shear flow $t_\mathrm{adj}$. We estimate the relative energy dissipation rate via
\begin{align}
\label{eq:dissipation_rate}
    \frac{\dot{E}}{E} = \frac{E(t_\mathrm{max}) - E(t_\mathrm{adj})}{E(t_\mathrm{adj})(t_\mathrm{max} - t_\mathrm{adj})},
\end{align}
where $t_\mathrm{max}$ is the vortex dissipation time of 10 orbits for the $64 \times 64$ simulation, and the simulation runtime of 100 orbits for the others.

The resulting dissipation rates are also shown in the left panel of  Fig.~\ref{fig:dissipation}. We fitted an exponential decay of form $\dot{E}/E = m \exp(-tN) + c$.  The dissipation rates are converging to a value of order $\dot{E}/E \sim 10^{-4} \Omega $. Given the dissipation rates in Fig.~\ref{fig:dissipation}, Eq.~\eqref{eq:dissipationrate_prediction} asserts an order of magnitude for the required under-saturation $\Delta q_\mathrm{req}$ to maintain the flow, i.e. $\Delta q_\mathrm{req} \sim \dot{E} / (\rho |\bm{v}| A C_\mathrm{E} L_\mathrm{s})$. Our seeded vortex has a total kinetic energy of order $E \sim 10^{-4}$ in code units, implying $\dot{E} \sim 10^{-8}$ for the fiducial simulation (see Fig.~\ref{fig:dissipation}). For $C_\mathrm{E} = 0.1$, $\rho \sim 1$,  a vortex area of $A \sim R_\mathrm{c}^2 = 10^{-2} H^2$, and velocities of order $|\bm{v}| \sim 10^{-2} c_\mathrm{s}$ therein, the required under-saturation that is to be sustained is of order $\Delta q\sim 10^{-4}$, which is modest.

Fig.~\ref{fig:dissipation} suggests that simulations with resolution less than $512 \times 512$ are not converged, and the observed vortex decay is mainly produced by numerical dissipation. This numerically-induced effective friction is therefore present in our fiducial $256 \times 256$-resolution simulations, and we proceed with the ansatz that the numerical dissipation in the calculations would be mirrored by an effective physical viscosity within a real protostellar disk.

Indeed, if the dissipation of a real disk vortex is driven by diffusion at the interface between the vortex flow and the Keplerian shear flow, then we expect the dissipation time, $t_\mathrm{d}$, to relate to the diffusion time. For a vortex of size, $s_\mathrm{v}$, and (numerical) kinematic viscosity $\nu$, this is given by
\begin{align}
\label{eq:disstime_prediction}
    t_\mathrm{d} = \frac{s_\mathrm{v}^2}{\nu} = \frac{1}{\alpha}\left(\frac{s_\mathrm{v}}{H}\right)^2 \frac{1}{\Omega},
\end{align}
where we have expressed $\nu$ in terms of the commonly used $\alpha$-parameterization \citep[][]{Shakura1973}
\begin{align}
\label{eq:alpha_def}
    \nu = \alpha c_\mathrm{s}H
\end{align}
Using $s_\mathrm{v} = R_\mathrm{c} = 0.1 H$, Eq.~\eqref{eq:disstime_prediction} implies an effective turbulence parameter of $\alpha \approx 4 \cdot 10^{-4}$, which is in line with the dissipation required to generate observed time scales for protostellar disk evolution \citep[e.g.,][]{Flock2017}. It is also consistent with the results of \citet[]{Godon1999}, who found that the exponential decay time, $\tau$, depends on the viscosity parameter $\alpha$. For $\alpha = 10^{-4}$, the anticyclonic vortices in \citet[]{Godon1999} propagate for $\tau \sim 50$ orbits, a duration comparable to that observed in our fiducial simulation with $256 \times 256$ resolution. 

\subsection{Vortex preservation and strengthening}

\label{sect:vortex_preservation}

\begin{figure*}[t]
    \centering
    \includegraphics[width = \textwidth]{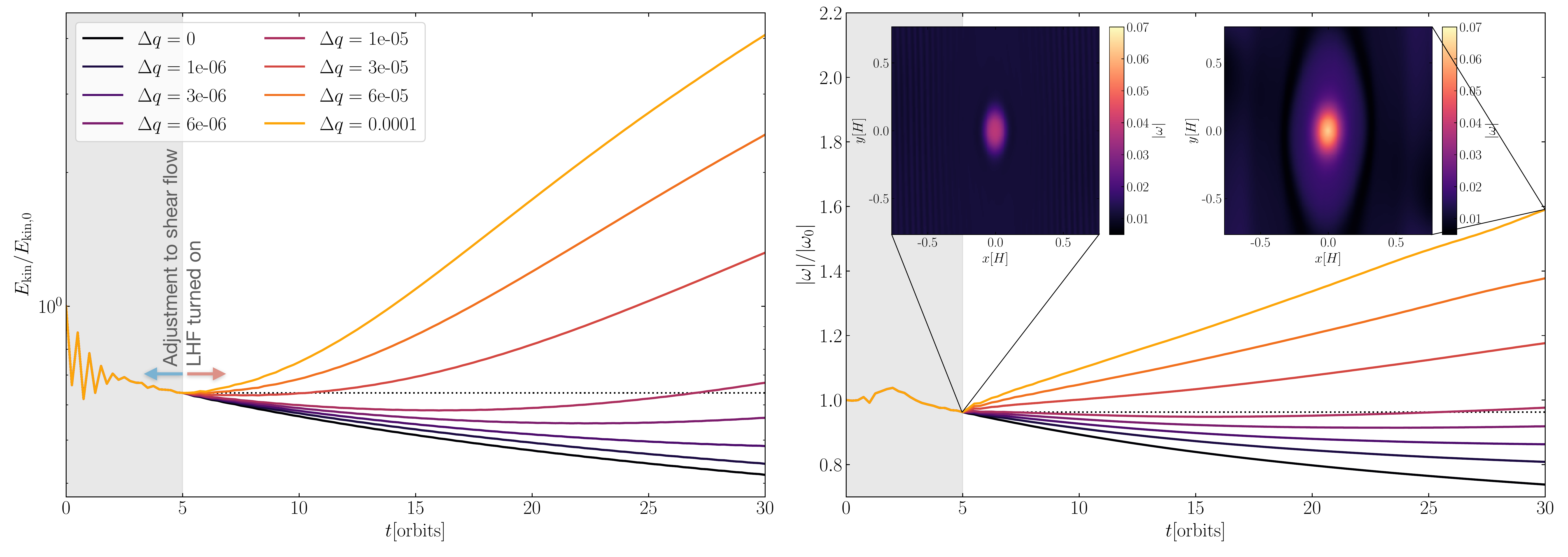}
    \caption{Like Fig.~\ref{fig:dissipation}, but for a fixed resolution of 256 $\times$ 256. Each color corresponds to a different under-saturation $\Delta q$ which drives the heat engine that is turned on after 5 orbits. The gray shaded region marks this period in which the vortex adjusts to the shear flow. The right panel includes two snapshots of the vorticity of the simulation with the strongest under-saturation. \new{Note that the vorticity colorbars have a different scaling to those in the right panel of Fig.~\ref{fig:dissipation}, reflecting the now much stronger vortices.}}
    \label{fig:preservation}
\end{figure*}

Fig.~\ref{fig:preservation} compares the fiducial simulation with simulations that incorporate heat engines of varying efficacy. Specifically we examine the influence of water vapor under-saturations in the range $10^{-6} \leq \Delta q \leq 10^{-4}$. Simulation run-times are 20 orbits. For simulations with $\Delta q \gtrsim 10^{-5}$, the vortex is maintained and both vorticity and vortex kinetic energy increase with time. The efficiency with which the vortex strengthens is somewhat less than what our order of magnitude estimate in Sect.~\ref{sect:vortex_dissipation} predicted. While vortices in the simulations with $\Delta q \lesssim 10^{-5}$ dissipate (albeit at a slower rate than in the reference simulation without heat engine) the vortices seeded into simulations with $\Delta q \gtrsim 10^{-5}$ increase both their vorticity and their size. An example of this growth can be seen for $\Delta q = 10^{-4}$ in the right panel of Fig.~\ref{fig:preservation}. \new{Increasing vorticity and vortex velocities also lead to a relative increase in viscous dissipation of vortical energy, which is why neither kinetic energy nor vorticity increase linear in $\Delta q$.}

We deliberately chose under-saturations $\Delta q$ to keep velocities sub-sonic within the run time of 20 orbits. This is in contrast to \citet[]{Les2015} who investigated vortices in the context of disk gaps. Their vortices are formed and maintained by the the Rossby Wave instability \citep[][]{Lovelace1999} and exhibit shock-inducing velocities, which generate dissipation. Our vortices would undergo a similar fate if they received energy input sufficient to drive super-sonic velocities. For example, the vortex in the simulation with  $\Delta q = 10^{-4}$ (yellow line in Fig.~\ref{fig:preservation}) increases its peak velocities at a rate of $7.5 \cdot 10^{-3} c_\mathrm{s}/\mathrm{orbit}$. If maintained, such boosting would lead to super-sonic velocities at a run time of about 120 orbits.

\subsection{Vortex formation via latent heat flux}
\label{sect:vortex_formation}

\begin{figure*}[t]
    \centering
    \includegraphics[width = \textwidth]{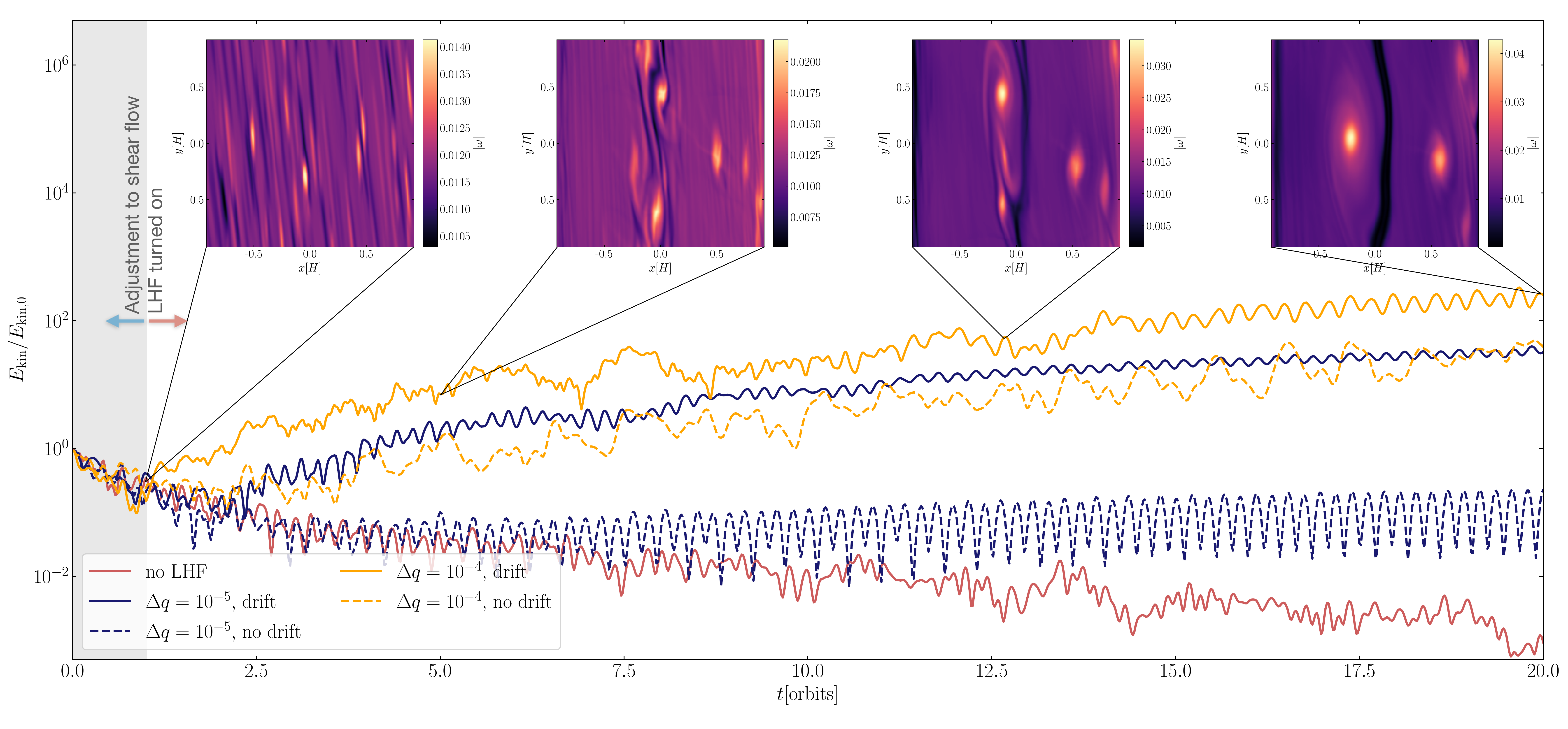}
    \caption{Evolution of the kinetic energy for random vortex initial condition for five different setups. Except for the run where LHF is turned off (red line), the adjustment time is set to one orbit (gray shaded region). Snapshots throughout the evolution of the absolute vorticity are shown for the simulation with $\Delta q = 10^{-4}$ and drift included.}
    \label{fig:vortex_cascade_energy}
\end{figure*}

Anticyclonic vortices are believed to form through a number of disc processes. Suggested mechanisms have included the Rossby wave instability \citep[][]{Lovelace1999} and the Vertical Shear Instability \citep[VSI, e.g.,][]{Nelson2013, Manger2018}. The latter phenomenon draws its energy from the vertical gradient in orbital frequency caused by radial variations in temperature and entropy. In this section, we explore the potential of the latent heat flux to generate vortices by tapping the thermodynamic disequilibrium between the gas flow and the icy layer in the presence of a background relative drift between the two layers. Note that because the latent heat flux is driven by a phase disequilibrium, and not by a temperature gradient, it is distinct from the VSI or the thermal instability investigated by \citet[][]{Owen2020}.

So far, we have treated the dust layer as a distinct laminar surface. In a more realistic disk, the coupling of dust and gas (characterized by the stopping time $t_\mathrm{s}$, a proxy for the particle size) leads to an equilibrium drift of particles both azimuthally and radially, even if $\bm{v} = 0$ in the Keplerian frame. For a pressure gradient
\begin{align}
\eta = -\frac{1}{2}\left(\frac{H}{R}\right)^2 \frac{\dif \ln \rho}{\dif \ln R},
\end{align}
and local dust-to-gas ratio, $\mu$, the equilibrium relative streaming velocity $\bm{w}$ in the Keplerian frame is given by \citep[][]{Nakagawa1986drift, Squire2018}
\begin{align}
\label{eq:nakagawadrift}
    \bm{w} = -2 \frac{\eta v_\mathrm{K}(1+\mu)\mathrm{St}}{(1+\mu)^2 + \mathrm{St}^2}\hat{\bm{x}} + \frac{\eta v_\mathrm{K}\mathrm{St}^2}{(1+\mu)^2 + \mathrm{St}^2}\hat{\bm{y}},
\end{align}
where $v_\mathrm{K} = R\Omega$ and $\mathrm{St} = t_\mathrm{s}\Omega$ is the dimensionless stopping time or Stokes number. Typically, $\mathrm{St} \sim 10^{-3} - 10^{-2} \ll 1$ \citep[see e.g.,][]{Testi2014}, and our subsequent discussion assumes that the stopping time is in this regime.

Vertical stratification of the particle mid-plane leads to height-dependent dust-to-gas ratio, $\mu$, and an attendant relative velocity $\bm{w}$ that decreases with height: The dust-rich mid-plane is less coupled to the gas flow and has a near-Keplerian orbital velocity in equilibrium. By contrast, the upper layers are more coupled to the gas flow and orbit close to the sub-Keplerian gas velocity of $(1-\eta)v_\mathrm{K}$.

This intrinsic vertical shear has long been thought to drive Kelvin Helmholtz Instability \citep[KHIs, ][]{Weidenschilling1980}, which mix the dust layer with the gas flow, and in the process prevent settling of the dust to a razor-thin layer. We envision particle mixing in the vortex to proceed in a similar manner, as particles in the dust-rich mid-plane likewise tend to collectively decouple from the gas flow even it deviates from the equilibrium sub-Keplerian stream lines. This decoupling becomes particularly prevalent for dust-rich disk regions with super-solar metallicities. There, the KHI becomes increasingly ineffective in turning over the now-massive particle stream, leading to a development of a high-density mid-plane cusp \citep[][]{Sekiya1998, Youdin2002, Gomez2005}. We discuss mixing in a disk subject to KHI further in Sect.~\ref{sect:mixing_model}.

The underlying equilibrium stream in Eq.~\eqref{eq:nakagawadrift} imposes asymmetry onto an anticyclonic vortex supported by the Keplerian shear.  The relative velocity between particles and vortical flow and thus the latent heat flux in Eq.~\eqref{eq:LHF_first} is decreased (increased) on the side towards (away from) the star and trailing (leading) the vortex's orbit, depending on if the Nakagawa drift is parallel (or anti-parallel) to direction of the vortex stream lines.

For the vortices simulated here, peak velocities are around 10\% of the sound speed. For fiducial values of $\mathrm{St} = 0.1, \mu = 0.02, \eta = 0.005$, and $v_\mathrm{K} = 0.05 \,c_\mathrm{s}$, the radial and azimuthal components of the background drift, $\bm{w}$ are of order 1\% and 0.1\% of the sound speed respectively. These deviations are an order of magnitude smaller than vortex flow velocities, and are thus of minor importance for the overall structure of the flow.

For weaker vortices, however, the background drift provides a significant contribution to the flow speed that drives mixing. In fact, even without a stable preexisting vortex, the background drift alone may be able to bootstrap a vortex flow, if the gas flow is sufficiently under-saturated. In order to investigate this we seeded a simulation with a random vortex field consisting of 100 vortices of with circulation $ 0.01 \leq \Gamma \leq 0.2$ and size $ 0.01 \leq R_\mathrm{c}/H \leq 0.1$, uniform-randomly distributed. We test this initial condition for setups with background drift both enabled and disabled, and compare to a reference simulation without latent heat flux. The background drift is implemented only insofar that the relative velocity in Eq.~\eqref{eq:nakagawadrift} is added to the relative velocity in Eq.\eqref{eq:LHF_first}. Hereby, we choose $\mathrm{St} = 0.1$ and $\mu = 0.02$ which correspond to marginally coupled dust particles and a typical dust-to-gas ratio \citep[e.g.,][]{Birnstiel2012}. We test two under-saturations of $\Delta q = 10^{-5}$ and $\Delta q = 10^{-4}$, both of which were enough to maintain the vortex in Fig.~\ref{fig:preservation}. For the two simulations that include the latent heat flux, the flux is enabled after a vortex adjustment time of one orbit. Simulations were run for a total duration of 20 orbits. 

Fig.~\ref{fig:vortex_cascade_energy} shows the evolution of the normalized kinetic energy. In the reference simulation where LHF is disabled, the initial condition dissipates and no lasting vortices survive (or form). Simulations with latent heat flux and drift (solid lines in Fig.~\ref{fig:vortex_cascade_energy}) are able to develop vortices from the initial condition. Fig.~\ref{fig:vortex_cascade_energy} includes four snapshots of the vorticity throughout the evolution of the $\Delta q = 10^{-4}$ simulation. At 1 orbit, when the latent heat flux is turned on, the initial vortex field has been sheared out by the Keplerian flow and a few anticyclonic vortices can be identified. Cyclonic vortices have been destroyed by the shear flow. The surviving vortices are subsequently strengthened by the latent heat flux, interact with each other, and ultimately merge, forming one large vortex spanning about a scale height in radial direction at 20 orbits. The maximum vorticity increases about a factor of three. The simulations with latent heat flux turned on follow the precedent set in Fig.~\ref{fig:preservation} where higher under-saturations, $\Delta q$, allow for stronger vortices. This is also seen in simulations without drift (dashed lines in Fig.~\ref{fig:vortex_cascade_energy}). Only the simulation with the greater latent heat flux is able to produce stable vortices, while the simulation with $\Delta q = 10^{-5}$ dissipates in a manner similar to that observed in the reference simulation.

\section{Discussion}
\label{sect:discussion}

Our analytical considerations and a variety of rudimentary numerical simulations point toward the possibility that hurricane-like vortices may exist in protoplanetary disks. In Sect.~\ref{sect:vortex_dissipation}, we measured dissipation rates of the vortex initial condition in \citet[][]{Adams1995} for a number of resolutions. In Sect.~\ref{sect:vortex_preservation}, we quantified the thermal disequilibrium between under-saturated gas flow and ice-coated dust required to maintain, or strengthen a dissipating vortex. As predicted in Eq.\eqref{eq:dissipationrate_prediction}, the required disequilibrium and the associated latent heat flux depends on the dissipation rate of the vortex. In Sect.~\ref{sect:vortex_formation}, we investigated the potential of the latent heat flux to form strong vortices from a random vortex initial condition.

To better understand the applicability of our work to real protoplanetary disks, it is thus worth discussing applicability of our model to real disks and the feasibility of latent heat fluxes required to balance expected vortex dissipation rates. We also discuss avenues for a mixing model, and the impact on dust trapping and planetesimal formation.

\subsection{Vortex streamlines in two and three dimensions}
\label{sect:3dmodel}

\begin{figure}
    \centering
    \includegraphics[width = \linewidth]{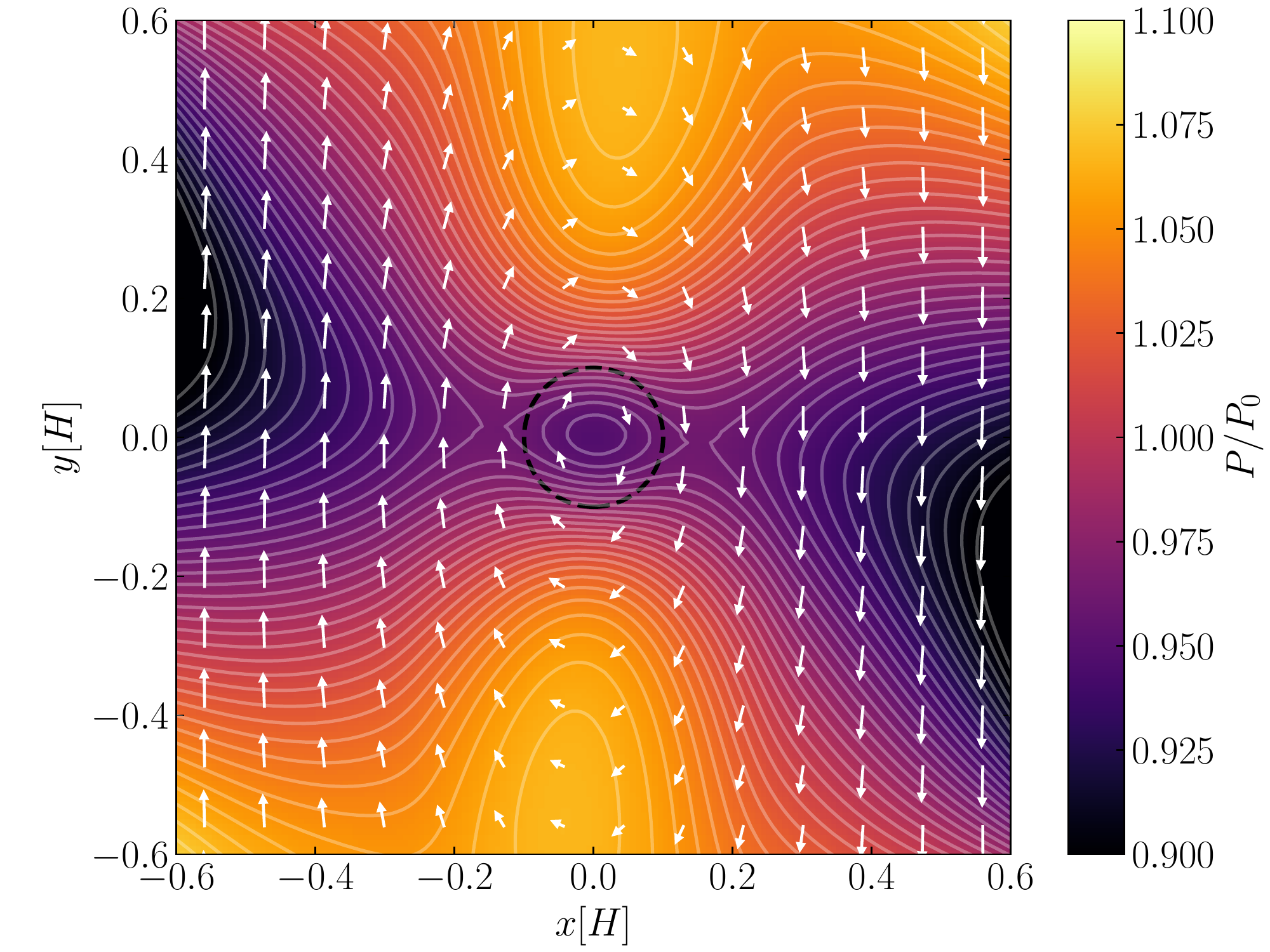}
    \caption{Final snapshot of the pressure of the $\Delta q = 10^{-4}$ simulation in Fig.~\ref{fig:preservation}, with contours representing isobars. Overlaid are the streamlines of the flow. The black circle marks the characteristic radius $R_\mathrm{c}$ with which the vortex was initialized.}
    \label{fig:vortex_final_streamlines}
\end{figure}

\begin{figure}
    \centering
     \includegraphics[width = \linewidth]{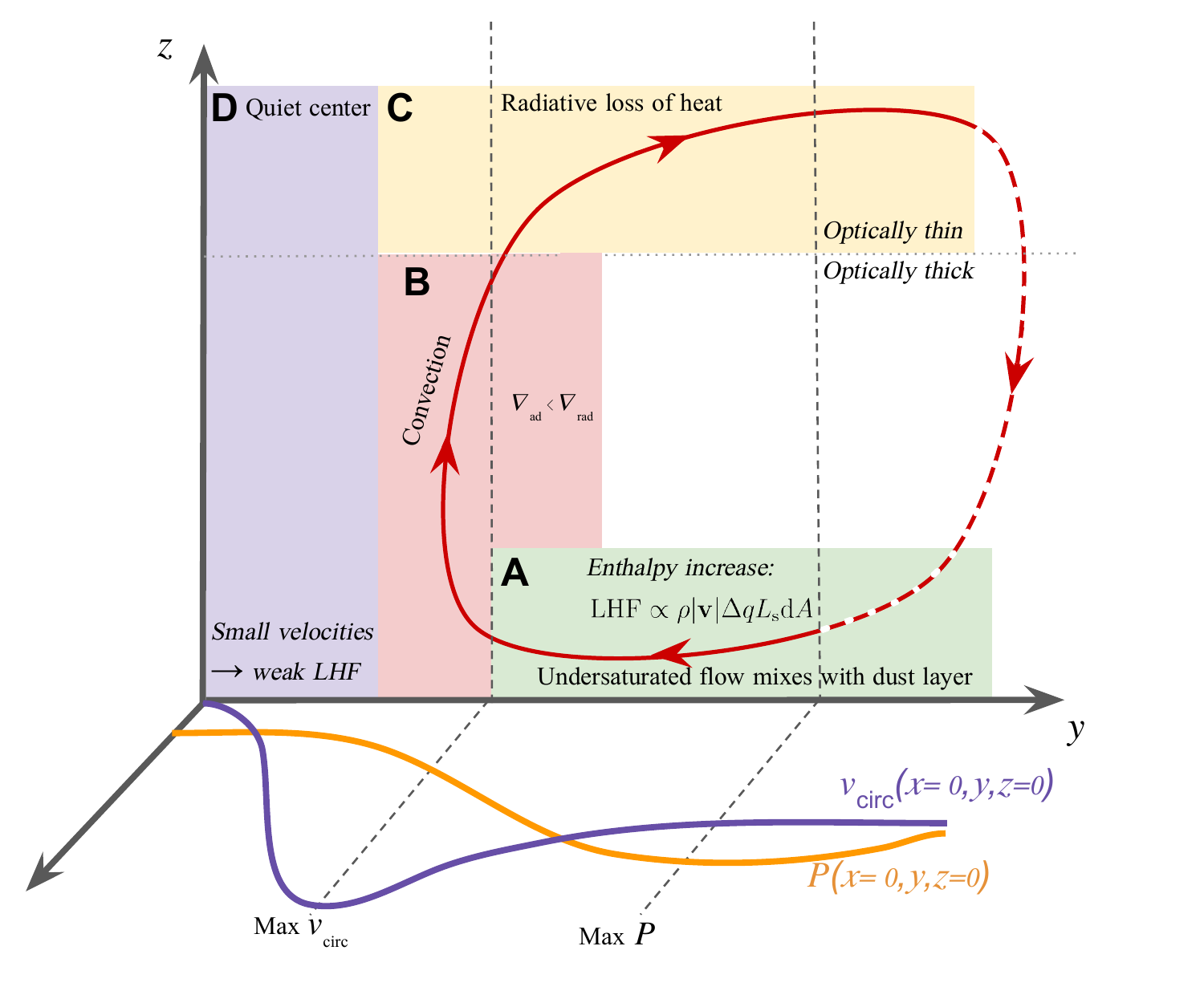}
    \caption{Proposed transverse circulation (top panel) in analogy to terrestrial hurricanes \citep[][]{Emanuel1991}. $y=0$ is the center of the vortex. The bottom part shows the run of pressure and circular velocities as seen in Fig.~\ref{fig:vortex_final_streamlines}.}
    \label{fig:vortex_3d_idea}
\end{figure}

We restricted our first-step numerical approach to two dimensions, yet hurricanes on Earth are characterized by three dimensional streamlines. In addition to the strong vortical flow, there is a transverse circulation. There, near-surface air streams radially inwards and is subject to the latent heat flux. The conversion of latent to sensible heat via rainfall drives vigorous upwards convection in the eyewall, which decreases the central pressure thus drawing in more ambient air. This strengthens the winds and thus also the latent heat flux, thus leading to the growth. In the eye wall, the warm air expanses upwards and cools until the temperature of the convective flow transitions smoothly into the ambient temperature profile. In order to mechanically maintain the flow, and allow under-saturated air into the system, there is a radial outflow at the tropopause, where excess heat is radiated away.  With this mechanism in mind, it is worthwhile considering plausible forms for three-dimensional streamlines that create analogous heat engines in protoplanetary disks. 

The key difference between the Earth case and the protoplanetary disk case is the strong Keplerian shear in the latter, which rapidly tears radially extended flow structures into azimuthally elongated bands, and also rapidly dissipates retrograde (cyclonic) vortices \citep[see e.g.,][]{Godon1999}. By contrast, prograde (anticyclonic) vortices are supported by the shear flow and dissipate significantly more slowly in the absence of an energy source. Moreover, there are a number of physical mechanisms that have long been posited to produce large-scale vortices (see Sect.~\ref{sect:vortex_formation}). The novelty of our approach derives from our focus on the applicability of the latent-heat flux driven heat engine to anticyclonic vortices. 

An anticyclonic vortex in geostrophic balance is a region of increased pressure relative to the ambient gas pressure in the disk. Fig.~\ref{fig:vortex_final_streamlines} shows a pressure map of the final snapshot of the $\Delta q = 10^{-4}$ simulation in Fig.~\ref{fig:preservation} with velociy flow field plotted on top. While pressure in the vortex center is indeed higher than the ambient pressure (required to maintain geostrophic balance), the pressure maximum occurs at around $(x,y) \sim (0,0.5 H)$, where circular velocities are high and the shear contribution is zero. It is here where we draw the analogy to a Hurricane on Earth and expect convection away from the mid-plane to mechanically maintain an additional transverse circulation.

We show a schematic of our proposed transverse circulation in  Fig.~\ref{fig:vortex_3d_idea}.  In analogy to the hurricane on Earth, the vortex flow picks up enthalpy of water sublimation by mixing with icy dust grains as gas parcels spiral towards the local minimum where azimuthal velocities are expected to be maximum (A). Note that we do not expect the inflow to proceed axisymmetric, but mainly azimuthally displaced from the vortex center as this where the pressure gradient $\partial P/\partial \varpi$ is maximal (see Fig.~\ref{fig:vortex_final_streamlines}). In order for the additional enthalpy to drive vertical convection (B) and maintain the pressure gradient, it must be energetically favorable for the additional heat to be carried dynamically rather than radiatively. For convective instability \citep[e.g.,][]{Lin1980}
\begin{align}
\label{eq:convective_crit_crude}
    \nabla_\mathrm{rad} > \nabla_\mathrm{ad} = \frac{\gamma - 1}{\gamma},
\end{align}
where the radiative temperature gradient is given by \citep[e.g.,][]{Kippenhahn1990}
\begin{align}
    \nabla_\mathrm{rad} = \frac{3}{16\pi acG}\frac{\kappa P}{T^4} \frac{l}{m},
\end{align}
where $a = 7.57 \cdot 10^{-15}$ erg cm$^{-3}$ K$^{-4}$ is the radiation constant, $\kappa = \kappa(T)$ the opacity, and $l/m$ the energy flux per unit mass. For convection, it must satisfy
\begin{align}
\label{eq:convective_crit}
    \frac{l}{m} > \frac{16\pi}{3}\frac{\gamma -1}{\gamma} acG \frac{T^4}{P\kappa(T)} \sim  10^{-3} \ \frac{\mathrm{erg}}{\mathrm{g \ s}},
\end{align}
where we have inserted $T = 150$ K, $P = 10$ erg/cm$^3$ (see Fig.~\ref{fig:sat_ratio}), and a fiducial mean opacity of order 1 cm$^2$/g corresponding to an optically thick mid-plane \citep[e.g.][]{Henning1996}. The latent heat flux per unit mass is of order $\mathrm{LHF}/m \sim L_\mathrm{s} C_E |\bm{v}|\Delta q / L$, where $L$ is a characteristic length scale over which the latent heat is distributed. For $C_\mathrm{E} = 0.1$, $|\bm{v}|/c_\mathrm{s} = 0.1 $, $\Delta q = 10^{-4}$, $L/H = 0.1$, and at $R = 3.6$ AU, our fiducial model yields $\mathrm{LHF}/m \sim 0.1$ erg/g/s. Given these rough estimates, this energy flux density would satisfy Eq.~\eqref{eq:convective_crit} by two orders of magnitude, and thus drive convection. Once the disk becomes optically thin, however, and $\kappa$ decreases, $\nabla_\mathrm{rad}$ increases and excess heat can be radiated away. The flow approaches thermodynamic equilibrium with the stellar radiation field, and loses the ability to drive convection ultimately leading to a radial outflow (C in Fig.~\ref{fig:vortex_3d_idea}), which mechanically maintains the structure. We expect the vortex center --- in analogy to the ''eye'' of the terrestrial hurricane --- to be relatively quiet, as rotational velocities, as well as corresponding latent heat fluxes, are minimal. Part A, B, and C in Fig.~\ref{fig:vortex_3d_idea} correspond to the first three legs of the Carnot cycle that characterizes a mature hurricane \citep[][]{Emanuel1991}.

Note that the convective instability we invoked to mechanically maintain the vortex can be produced by a variety of mechanisms in protoplanetary disks. Accretion, for example driven by the magneto-rotational instability \citep[][]{Balbus1991}, can also render optically thick disks convective \citep[][]{Garaud2007, Jankovic2021}, which would relax the additional requirement that Eq.~\eqref{eq:convective_crit_crude} poses on the latent heat flux.

\subsection{Mixing model}
\label{sect:mixing_model}

In our parameterized model of the latent heat flux in Eq.~\eqref{eq:LHF_first}, the mixing efficiency is described by the turbulent exchange coefficient $C_\mathrm{E}$, which describes the efficiency with which water vapor is mixed into the flow thereby enabling it to transfer its latent heat. Past experimental measurements of $C_\mathrm{E}$ \citep[e.g.,][]{Verma1978} while difficult to transfer to the protoplanetary disk context, have been accompanied by theoretical considerations in the Earth context. In this section, we will build upon this framework to calculate an order-of-magnitude analytical prediction of $C_\mathrm{E}$ in the protoplanetary disk context, thereby motivating our choice of $C_\mathrm{E} = 0.1$.

For a latent heat flux from Earth's ocean to the atmosphere, the latent heat transfer coefficient above the interface is known to depend on wind speed and sea roughness, i.e. wave height, period, steepness, as well as the relative orientation of the wind velocity vector to the direction of wave propagation \citep[e.g.,][]{Cronin2019}. In addition, the transfer efficiency is expected to decrease with height above the boundary layer as turbulent eddies cascade to decay in the presence of gas viscosity. This decrease is typically modeled using a semi-empirical logarithmic law, and modified by functions $\psi$ developed in Monin-Obukhov Similarity (MOS) theory \citep[][]{Monin1954}, which handle complexities stabilizing (stratifying) or destabilizing (convective) conditions. The wind speed at height $z$ above the boundary layer then can be written as
\begin{align}
\label{eq:log_wind_profile}
    v(z) = \frac{v_\mathrm{f}}{\kappa_\mathrm{f}} \left[\ln\left(\frac{z}{z_0}\right)+ \psi(z,z_0)\right],
\end{align}
where $\kappa_\mathrm{f} \approx 0.4$ is the von K\'{a}rm\'{a}n constant, $z_0$ is the interface's roughness length, and $v_\mathrm{f}$ the friction velocity which in turn depends on shear stress $\tau$ and fluid density $\rho$ via $v_\mathrm{f}^2 = \tau/\rho$. Given a roughness length $z_0$ and MOS correction term, the turbulent exchange coefficient for latent heat can be expressed in the form \citep[see][]{Garratt1994}
\begin{align}
\label{eq:exchange_coeff_def}
    C_\mathrm{E} = \left(\frac{\kappa_\mathrm{f}}{\ln(z/z_0) - \psi(z,z_0)}\right)^2.
\end{align}

In a protoplanetary disk, the boundary between dust grains and gas fluid is not distinct as it is with the ocean-air interface on Earth. Instead, the pressure gradient, $\eta$, prescribes the intrinsic scale height of the particle layer as regulated by KHI \citep[e.g.,][]{Chiang2008, Gerbig2020}. As a consequence, even in equilibrium, gas and dust are not well-separated flows, but form a combined fluid, with a vertical gradient in dust-to-gas ratio and resulting drift velocities. Our goal is to use the established theory of KHI in disks to construct a function $\psi(z,z_0)$ that satisfies the semi-logarithmic wind profile in Eq.~\eqref{eq:log_wind_profile} allowing us to calculate an exchange coefficient via Eq.~\eqref{eq:exchange_coeff_def}. 

A flow with velocity component $v(z) = (v_\phi^2 + v_R^2)^{1/2} $ perpendicular to a gravitational acceleration $\bm{g} = g\bm{\hat{z}}$ is subject to KHI if its Richardson number 
\begin{align}
\label{eq:Richardson_number}
    \mathrm{Ri} = \frac{(g/\rho)(\partial\rho/\partial z)}{(\partial v/\partial z)^2}\, ,
\end{align}
falls below a critical number, for Cartesian flows given by $\mathrm{Ri} = 1/4$ \citep[][]{Chandrasekhar1961}. In equilibrium, and for well coupled particles with $\mathrm{St} < 1$, Eq.~\eqref{eq:nakagawadrift} implies for the azimuthal flow velocity
\begin{align}
\label{eq:nakagawa_azimuthal}
 v_0(z) = v_\mathrm{K} \left(1- \frac{\eta}{1 +\mu(z)}\right),
\end{align}
such that in absence of a dust, i.e. far away from the mid-plane, the gas moves on a sub-keplerian orbit. With this velocity profile, and $\mathrm{Ri} = \mathrm{const}$, Eq.\eqref{eq:Richardson_number} can be integrated leading to a profile for the dust-to-gas ratio $\mu$ given by \citep[][]{Chiang2008}
\begin{align}
    \mu(z) &= \sqrt{\frac{1}{f(z)}} -1, \\
    f(z) &\equiv \frac{1}{(1+\mu_0)^2}+\frac{1}{\mathrm{Ri}}\left(\frac{z}{\eta R}\right)^2,
\end{align}
where $\mu_0$ is a integration constant corresponding to the mid-plane dust-to-gas ratio. We assume that the dust-rich mid-plane orbits at an approximately Keplerian rate --- valid if the particle concentration is high enough to develop an over-dense mid-plane cusp with $\mu_0 \gtrsim 1$ \citep[][]{Sekiya1998, Youdin2002, Gomez2005} --- and write the ``wind'' profile as
\begin{align}
\label{eq:windprofile}
    v(z) = v_0 - v_\mathrm{K} = -  \Omega \eta R f^{\frac{1}{2}}.
\end{align}

The shear stress $\tau$ can be defined as
\begin{align}
    \tau(z) = \rho \nu \frac{\partial v}{\partial z} = \rho c_\mathrm{s}^2 \frac{\alpha}{\mathrm{Ri}}\frac{ z}{\eta R}f^{-\frac{1}{2}},
\end{align}
where we used the $\alpha$-prescription in Eq.\eqref{eq:alpha_def}, yielding a friction velocity
\begin{align}
\label{eq:frictionvelocity}
    v_\mathrm{f} =  \sqrt{\frac{\tau}{\rho}} = c_\mathrm{s} \left(\frac{\alpha}{\mathrm{Ri}}\frac{z}{\eta R}\right)^{\frac{1}{2}}f^{-\frac{1}{4}}.
\end{align}
Inserting Eq.\eqref{eq:windprofile} and Eq.\eqref{eq:frictionvelocity} into Eq.\eqref{eq:log_wind_profile} yields the correction term $\psi(z,z_0)$, i.e.
\begin{align}
    \psi(z,z_0) = \ln\left(\frac{z}{z_0}\right) - \eta \kappa_\mathrm{f} \frac{R}{H}f^{\frac{1}{4}}\left({\frac{\mathrm{Ri}}{\alpha} \frac{\eta R}{z}}\right)^{\frac{1}{2}}.
\end{align}
The turbulent exchange coefficient in Eq.~\eqref{eq:exchange_coeff_def} is thus
\begin{align}
\begin{split}
    C_\mathrm{E}(z) = & \left[\vphantom{{\frac{1}{2}}^\frac{1}{2}}\right.\frac{2}{\kappa_\mathrm{f}}\ln\left(\frac{z}{z_0}\right) - \frac{\eta R}{H}\left({\frac{\mathrm{Ri}}{\alpha} \frac{\eta R}{z}}\right)^{\frac{1}{2}} \\ & \cdot \left(\frac{1}{(1+\mu_0)^2}+\frac{1}{\mathrm{Ri}}\left(\frac{z}{\eta R}\right)^2\right)^{\frac{1}{4}} \left.\vphantom{{\frac{1}{2}}^\frac{1}{2}}\right]^{-2}.
\end{split}
\end{align}
For the purposes of our estimate, we evaluate the exchange coefficient $C_\mathrm{E}$ at $z = \eta R$, which is a typical length scale for the particle layer \citep[e.g.,][]{Gerbig2020}, and notice that for Richardson number of order unity \citep[][]{Johansen2006KHI}, $f^{1/4}$ too is of order unity. Thus,
\begin{align}
\label{eq:exchange_coeff_final}
    C_\mathrm{E}(z = \eta R) \approx \left[\frac{2}{\kappa_\mathrm{f}}\ln\left(\frac{\eta R}{z_0}\right) - \frac{\eta R}{H}\left({\frac{\mathrm{Ri}}{\alpha}}\right)^{\frac{1}{2}}\right]^{-2}.
\end{align}
In this form, the (equilibrium) turbulent exchange coefficient depends on the local disk properties pressure gradient $\eta$, aspect ratio $H/R$, viscosity $\alpha$, the flow Richardson number $\mathrm{Ri}$, and the ratio of roughness scale $z_0$ to characteristic length scale $\eta R$. Since we did not consider potential vortex modifications to the velocity profile but used the equilibrium velocity profile $v_0$, Eq.~\eqref{eq:exchange_coeff_final} likewise is a prescription for the equilibrium exchange coefficient.

\begin{figure}
    \centering
    \includegraphics[width = \linewidth]{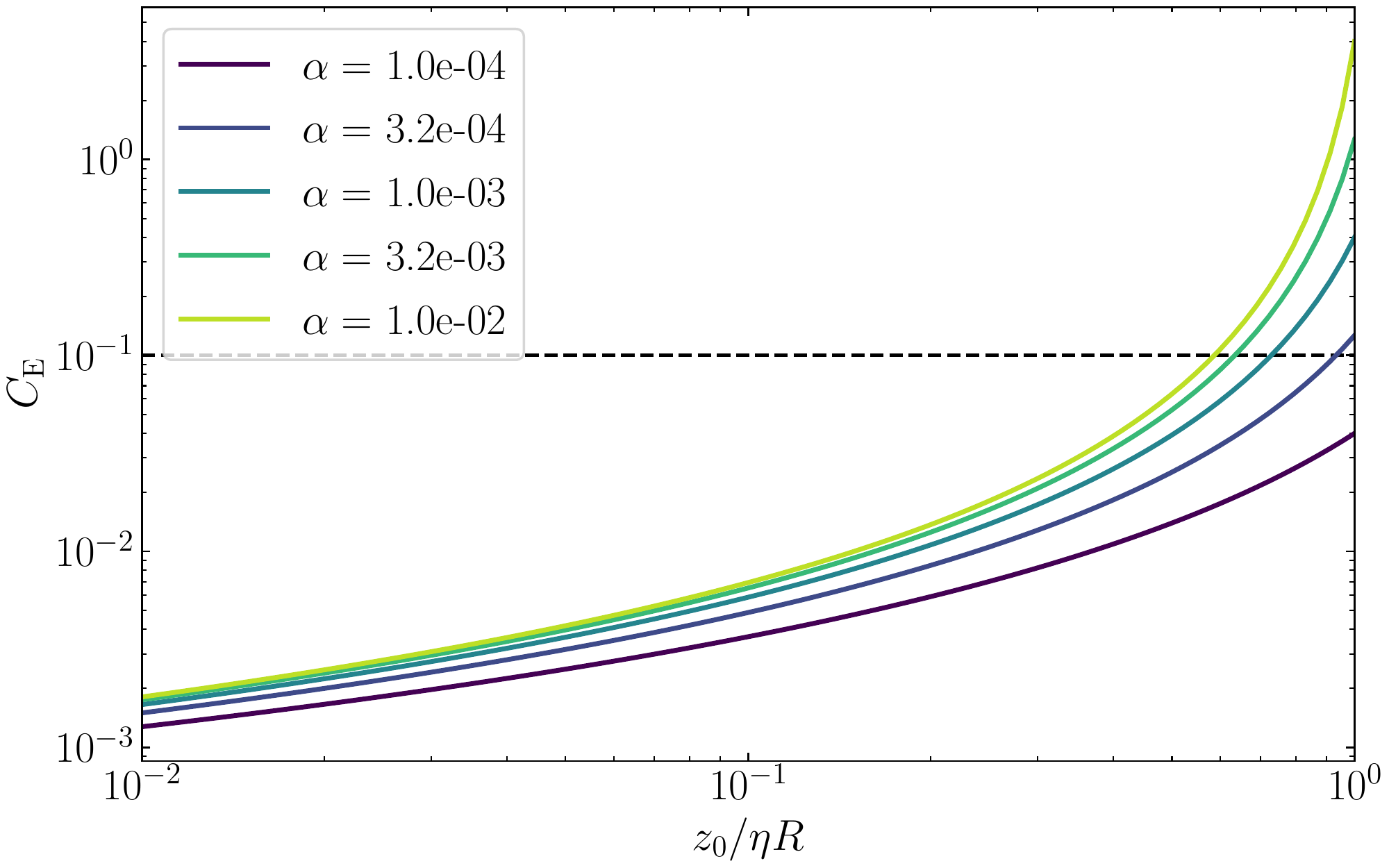}
    \caption{Turbulent exchange coefficient in Eq.~\eqref{eq:exchange_coeff_final} vs roughness length $z_0$ for different $\alpha$ values. The dashed line represents our choice of $C_\mathrm{E} = 0.1$. We assumed $\eta = 5\cdot 10^{-3}$, $H/R = 0.1$, and $\mathrm{Ri} = 1$.}
    \label{fig:exchange_coeff}
\end{figure}
However, we can use the roughness length $z_0$ to incorporate the effect of faster velocities, as faster wind speeds are expected to increase the roughness length. Fig.~\ref{fig:exchange_coeff} shows the turbulent exchange coefficient at a height $z = \eta R$ above the mid-plane vs roughness length $z_0$ for multiple values of $\alpha$. Our choice of $C_\mathrm{E} = 0.1$ is achieved for high $z_0$ (of order $\eta R/2$) and moderate $\alpha$-values. \new{Note that while high $\alpha$-values lead to more efficient mixing, and thus to an increase in latent heat flux and energy input, they also would cause the vortex to decay at a faster rate \citep[compare to e.g.,][]{Godon1999}. The overall impact of an external $\alpha$ is therefore not obvious, and will be studied in more detail in future work.}

\subsection{Dust trapping and planetesimal formation}

Anticyclonic vortices are associated with maxima in gas pressure and are thus commonly invoked as a dust trap \citep[e.g.][]{Heng2010, Raettig2015, Owen2017} that can break the meter barrier and lead to planetesimal formation \citep[see][for a review]{Chiang2010}. In models of planetesimal formation that rely on such a dust trapping mechanism, the lifetime of traps can be a bottleneck \citep[e.g.,][]{Lenz2019, Gerbig2019}, since vortices in accretion disks have finite lifetimes in the absence of any support mechanisms \citep[][]{Lesur2009}

Therefore, if the heat engine successfully operates in disks and elongates vortex life times, we expect the framework discussed here to provide a zeroth-order favoring of planetesimal formation. Planetesimal formation feedback, however, would remove dust grains from the disk, in the process decreasing the surface available for water vapor sublimation and water ice deposition. This may shut off the heat engine, much as a terrestrial Hurricane is shut off when it hits land. In addition, if dust feedback is considered, the inertia of the captured dust stream can also destroy a vortex \citep[][]{Fu2014}. Similarly, the  convection we invoked to maintain the heat engine may displace particles vertically, which can disrupt regions of gravitational collapse, similar to the simulations studied in \citet[][]{Schaefer2020, Gole2020}, where gas turbulence hinders sedimentation.
Another point of consideration is that the phase disequilibrium driving the vortex requires that vapor and water ice co-exist in the immediate environment of the vortex. This constrains the ideal location to just outside the ice line. \new{Promisingly, \citet[][]{Pfeil2021} found that vortices can be found throughout the disk, including the regions around the iceline. On the other hand, } in global simulations by \citet[][]{Flock2020}, a large scale vortex forms at 30 AU. Such a structure would be wholly unaffected by latent heat flux. 

The overall impact to planetesimal formation in the context of vortices as dust trap location is therefore not obvious, and must be investigated in more detail --- ideally with numerical simulation that include dust particles.

\subsection{Conclusions}

Terrestrial hurricanes have been studied intensively (in part as a consequence of their fearsome economic importance \new{\citep[e.g.,][]{Emanuel2005}}) and they constitute truly remarkable self-organized systems powered by a flux of latent heat and maintained by the thermodynamic disequilibrium between the ocean surface and the atmosphere. Moreover, it seems doubtful that even the existence of these cyclonic storms would have been proposed in the absence of direct observations and on the basis of purely theoretical investigations of idealized models of Earth's ocean-atmospheric structure. It is thus of interest to use terrestrial hurricanes as a guide to explore whether a similar situation can arise in protoplanetary disks. Our conclusion is if disk gas flows are under-saturated, particularly just outside the water ice line, then the essence of the hurricane mechanism may spring into operation.

Under-saturation incites sublimation of water from ice-coated dust-grains in the mid-plane, leading to an energy increase in the flow. Our heuristic two-dimensional simulations suggest that modest under-saturations are sufficient to maintain pre-existing vortices that would otherwise dissipate (Sect.~\ref{sect:vortex_preservation}). Our results also suggest that the shear inherent to gas and particle streams in protoplanetary disks can enhance the enthalpy transfer mechanism, and potentially cause formation of vortices  (Sect.~\ref{sect:vortex_formation}).

Three-dimensional simulations, ideally including the dust-particles independently,  are necessary to confirm the feasibility of hurrican-like storms in prostellar disks. As with hurricanes on Earth, a carnot engine in a disk will be inherently three-dimensional. Our work here merely suggests that latent heat fluxes can strongly affect the flow dynamics near the water ice line. Further investigation thus seems warranted.

\acknowledgments \new{The authors thank the referee for useful comments that significantly improved the manuscript.} GL acknowledges generous support from the Heising-Simons Foundation through Grant 2021-2802 to Yale University. KG thanks \new{Diana Powell, Thomas Pfeil, and} the Yale Exoplanet Seminar for useful discussions.

\software
{Athena++ \citep[][]{Stone2020}, Matplotlib \citep[][]{Hunter2007}, Numpy \citep[][]{Walt2011}, Scipy \citep[][]{Jones2001}} 
\newpage
\bibliography{references}{}
\bibliographystyle{aasjournal}

\end{document}